\documentclass[aps,pra,preprint,showpacs,superscriptaddress,groupedaddress,a4paper,longbibliography]{revtex4-1}
\usepackage{amsmath}
\usepackage{epsfig}
\usepackage{graphicx}
\usepackage{color,float}
\usepackage{amssymb}
\usepackage{epstopdf}
\usepackage[colorlinks=True, linkcolor=blue, citecolor=blue]{hyperref}
\usepackage{xcolor}
\begin{document}
\title{Emergence of N-body tunable interactions in \\ universal few-atom systems}
\author{M. T. Yamashita$^{1}$}\thanks{marcelo.yamashita@unesp.br}
\author{T. Frederico$^{2}$}\thanks{tobias@ita.br}
\author{Lauro Tomio$^{1}$}\thanks{lauro.tomio@unesp.br}
\affiliation{
$^{1}$Instituto de F\'isica Te\'orica, UNESP, 01140-700, S\~{a}o Paulo, Brazil.\\
$^{2}$Instituto Tecnol\'ogico de Aeron\'autica, 12228-900, S\~ao Jos\'e dos Campos, Brazil.\\
}
\begin{abstract}
A three-atom molecule AAB, formed by two identical bosons A and a distinct one B, is studied by considering 
coupled channels close to a Feshbach resonance. It is assumed that the subsystems AB and AA have, respectively, 
one and two channels, where, in this case, AA has open and closed channels separated by an energy gap.
The induced three-body interaction appearing in the single channel description is derived using the Feshbach 
projection operators for the open and closed channels. An effective three-body interaction is revealed 
in the limit where the trap setup is tuned to vanishing scattering lengths .
The corresponding homogeneous coupled Faddeev integral equations are derived in the unitarity limit. The s-wave 
transition matrix for the AA subsystem is obtained with a zero-range potential by a subtractive renormalization scheme 
with the introduction of two finite parameters, besides the energy gap. The effect of  the coupling between the 
channels in the coupled equations is identified with the energy gap, which essentially provides an ultraviolet 
scale that competes with the van der Waals radius - this sets the short-range physics of the system in the open channel. 
The competition occurring at short distances exemplifies the violation of  the ``van der Waals universality" 
for narrow Feshbach resonances in cold atomic setups. In this sense, the active role of the energy gap drives the 
short-range three-body physics.
\end{abstract}
\maketitle
\date{}
\section{Introduction - an overview of recent problems involving universality in few-body systems}
\label{intro}

The existence of several nuclear potentials with many free parameters, usually set to reproduce scattering
observables, raises the question about the possibility to make something relevant with a zero-range 
potential. From the absolute point-of-view it is not possible to describe the energy spectrum or 
the nuclei structure with a such quasi-no-parameter potential. However, the description of 
physical observables for specific nuclear systems, after identifying the relevant scales, is completely 
suitable for a Dirac-delta potential: this is in the core of the universality 
concept~\cite{2000aDelfino,2006braaten,2011Frederico}.

When applied to few-body systems, universality means, {\it grosso modo}, independence on details of the 
short-range part of the interactions which are describing such systems.
These systems appear in several few-body areas, being characterized for having their typical sizes, represented 
by the absolute value of the two-body scattering length $a_0$ and potential range $r_0$, such that the first is much 
greater that the second, $|a|\gg r_0$ - this property defines a weakly-bound system. There are 
many molecules in the atomic context, which satisfy the relation $|a|/r_0\gg1$~\cite{2010Chin}. 
In the nuclear context, these weakly-bound structures are well represented by halo nuclei, in which
we have one or more halo nucleons weakly-bound to a core nucleus.~\cite{fredericoppnp2012}.

Universality in few-body physics provided explanation to many interesting phenomena, which have been
extensively studied in recent years. The remarkable one is the Efimov effect~\cite{1970efimov}, which 
was stablished by studying bosonic-like three-body system with at least two subsystems having 
infinite scattering length. This effect became a paradigm when mentioning universal aspects of 
few-body systems - nowadays, the study of the relevant physical scales close to the unitary limit is 
known as ``Efimov physics''.

A natural three-body system with two-body subsystems exactly at zero energy does not exist. One of 
the systems found in nature that approaches the ideal situation is the helium trimer. It was suggested that
the Efimov effect could be verified in this molecule~\cite{1984CG}. The lack of any other natural 
three-body system close to the Efimov limit was the main reason why 
the original article from Efimov was considered for thirty years uniquely as a theoretical allegory. 
However, with the recent modern techniques involving ultracold traps, it was possible to 
measure the ground and first excited states of the helium trimer~\cite{2015-kunitski}, confirming 
the original suggestion and many other theoretical studies from different groups~\cite{1985Huber,1996Esry,1997Kolganova,2001Motovilov,2004Kolganova,2000Delfino,2000Gianturco,2002Gianturco,2012Roudnev,2014Wu,2017Kolganova,2018Kolganova}. 

The confirmation of Efimov states is also possible by producing them artificially. Some experiments with cold-atom
systems were able to observe signals of Efimov resonant levels by manipulating the two-body interactions.
The possibility to freely tune the two-body interaction in ultracold atomic traps brought 
unprecedented possibilities to artificially create the ideal condition for Efimov states~\cite{2006Kraemer}. 
The use of Feshbach resonances technique~\cite{1958Feshbach} to alter the two-body 
energies~\cite{1998Inouye,1998Courteille,1999Timmermans} in atomic traps produced a 
rich playground with plenty of possibilities to investigate the correlations between the physical scales.
Experimentalists are now able to continuously vary $a_0$ from large positive to large negative values, 
across the threshold at which a bound state turns into a resonance or a virtual state (see e.g. 
Ref.~\cite{fredericoppnp2012}). 

The continuous change of spatial dimension is also another interesting aspect~\cite{2019Garrido} that can now 
be explored in actual experimental realizations~\cite{2017Klauss}. The study of one, two or 
three-dimensional solitons has already start\-ed long ago (See \cite{2000Gammal,2005Abdullaev} and references therein). 
However, it still missing systematic experimental studies about the effect of the dimensional change in 
few-body observables, e.g., the Efimov effect is drastically affected by the spatial dimensional reduction. 
The disappearance of Efimov effect in fractional dimensions between three and two dimensions, including another few-body 
aspects involving a continuous change of spatial dimensions, has recently being investigated by 
some groups~\cite{2018Sandoval,2018aRosa,2018bRosa,2018Christensen,2018Abhishek,2019Yama,2019Beane,2019Garrido}. 
 
All the experimental possibilities regarding the change of dimensionality and atomic 
interactions are not possible in the nuclear context where the nucleon-nucleon interaction 
is rigid. A conjecture was raised in Ref.~\cite{2008Higa} that a resonant Hoyle state in $^{12}$C could 
emerge from an Efimov bound state. In order to investigate this conjecture, a recent work 
studied the screening of the Coulomb potential in three alphas~\cite{2020Phyu}.

The strong repulsion at short distances coming from the Coulomb potential is what 
prevents the appearance of Efimov effect in three alphas. However, in hot dense plasmas the 
electrons can, in principle, shield the repulsion coming from the protons favoring the formation 
of the Hoyle resonance. A continuous increase of the electronic density can also produce a similar 
effect, occurring in atomic traps, that is the change of the effective two-body interaction of the 
system. However, recent calculations made in Ref.~\cite{2020Phyu} exclude the possibility of a 
Hoyle state emerging from an Efimov bound state.

Another interesting aspect in the context of few-atom systems, which is the central point 
of this article, is that, at least theoretically, it is possible that induced few-atom (beyond two) 
forces appear~\cite{YamaEPL2006} when the cold atomic trap set-up is tuned near a narrow 
Feshbach resonance, where the closed channel is strongly coupled to the open 
channel~\cite{WangNPHYS2014}. If the control of only two-body interactions~\cite{1999Timmermans} 
revealed a wide horizon to study correlations among their observables, due to the change of  few-body 
scales~\cite{2020Paula}, the possibility to control three and more atomic interaction would give 
an unprecedented freedom to few-body community. 

Formally, the origin of the induced few-particle effective interactions in a single channel representation 
of  few-atom system goes back to the Feshbach decomposition of the Hilbert space in open channels ($P$-space)  
and closed channels ($Q$-space) , given that $P+Q=1$(details on the coupled-channel formalism applied to 
potential scattering can be found in the textbook by Canto and Hussein in Ref.~\cite{Canto2013}).
The $Q$-space represents the state where the two atoms interacts in a region of the potential where the scattering states
are closed. The $P$-space corresponds to potential well lower in energy where the scattering states are open. 
Actually, $P$ and $Q$ are associated with different spin states of the low partial waves of the 
atom-atom system (see e.g.  \cite{WangNPHYS2014}). In such framework,
attractive effective few-body forces arise from connected diagrams with the intermediate virtual propagation 
of the system in the $Q$-space as illustrated in~\cite{YamaEPL2006}. The strength is enhanced for narrow 
resonances, as the coupling between open and closed channels is larger in this case~\cite{WangNPHYS2014}.

The experimental evidences suggesting the possibility of few-atom forces come from the so called 
``van der Waals  universality"~\cite{BerPRL2011,ChinArXiv2011,WangPRL2012,SchmidtEPJB2012,NaidonPRL2014,NaidonPRA2014} 
of Efimov states across broad and narrow Feshbach resonances~\cite{2014chin,2017chin} with Lithium-Caesium  
 ($^6$Li-$^{133}$Cs) mixtures.  The ``van der Waals  universality" associates the position $(a_0<0)$ of the resonant 
 three-atom  recombination peak, originated when an Efimov state dives into the continuum, with van der Waals radius 
 $(\ell_{vdW})$, such that the ratio $a_0/\ell_{vdW}\sim -9$ is verified for broad resonances~\cite{BerPRL2011,2017chin}. 
However, the experimental results from Ref.~\cite{2017chin} indicated a dependence of the position of the Efimov 
resonance on the Feshbach resonance strength, deviating from the prediction of the single channel ``van der Waals" 
universality. Such observation suggests that close to narrow Feshbach resonances a single channel description 
may be poor, and beyond the expected large variations of the atom-atom scattering length, the three-body 
scale can also change, as also supported by the observation of resonant recombinations in the 
$^6$Li-$^{133}$Cs$_2$ experiments~\cite{2017chin}. 

Note that the length scale associated with the position of the triatomic recombination resonance is observed to be 
larger than the van der Waals length for narrow resonances~\cite{2017chin}. This was interpreted in Ref. \cite{2020Paula} 
as  a manifestation of the attraction due to the induced three-atom interaction, which dislocates the effective repulsive 
barrier~\cite{jonsell}, where the triatomic continuum resonance~\cite{2004Bringas} is formed, to distances larger than 
$\ell_{vdW}$. Therefore, the position of the narrow recombination resonance appears dislocated towards larger absolute 
values of the scattering length with respect to broad Feshbach resonances, as experimentally observed~\cite{2017chin}.

One should expect that, in few-atom systems driven by the Feshbach resonance mechanism, which can induce from 
three up to N-body interactions, systems beyond the atom-atom scattering length can be manipulated by tuning also 
the short-range scales associated with three, four and more particles~\cite{YamaEPL2006,2011hadizadeh,2019bazak,2020Paula}. 
Such exciting possibilities motivated us to explore the ideas we have sketched in Ref.~\cite{YamaEPL2006}. We write here  
the formalism of the three-body Faddeev equations for the bound state in the presence of open and closed channels. 
Following that, a practical framework can be formulated by identifying the relevant parameters of the Feshbach resonance, 
which control the induced three-body interaction and the associated short-range scale.

In this contribution, the AAB atomic bound state problem with coupled open-closed channels close to a  Feshbach resonance 
is formulated through the Faddeev equations for the wave function. The effective three-body force appearing in the single 
channel description is detailed by using the Feshbach projection operators in the open and closed channels. We discuss the 
interesting case where the scattering lengths vanish, such that in the open channel the direct interaction vanishes, while we 
show  formally that the Faddeev components of the wave function in the open channel survives, allowing the system to effectively 
interact in a single channel description through the coupling of the open channel with the closed one. Furthermore, the Faddeev 
formalism is derived explicitly for a zero-range model in one particular example of three-particle AAB system, where the AA 
subsystem has open and closed channels, by formulating in practice the ideas proposed in Ref.~\cite{YamaEPL2006}. 
The dependence of the short-range physics on the parameters of the Feshbach resonance in the 
AA subsystem is obtained with the corresponding meaning being explored qualitatively.

This work is organized as follows. In the next Sect.~\ref{sec2}, the Faddeev equation for the
bound state wave function for a general three-body model with open-closed channels is formulated. 
It is also discussed the single channel reduction and induced three-body interaction in the open channel. 
In addition, the effective three-body interaction appearing when the system has vanishing scattering length is
derived within the closed channel description of the three-body bound state. In Sect.~\ref{sec:AAB}, 
the AAB Faddeev equations for the bound state within a zero-range model with open and closed channels is derived in 
the case that the AA bosonic subsystem has one open and one closed channels, and by considering the AB interaction 
acting only in a single channel. Once obtained the two-body T-matrix for the AA subsystem within the zero-range interaction,
the integral equations for the AAB coupled channel model in the unitarity limit $(a\to\pm\infty)$ are derived, 
by considering the dependence on the Feshbach resonance parameters. In Sect. \ref{sec:final}, 
we present our final considerations.

\section{Open-closed channels three-body model}\label{sec2}

The open-closed channels model for three-atom interactions has altogether eight channels, 
as each pair can interact in an open or closed channel. 
The notation has to reflect such different physical situations, with an index $\alpha$ or $\beta$ 
for the corresponding channel wave function, in which the atom pairs can be in open or closed channels, 
being indicated by $\alpha_{ij}$, which runs over the two-body channels. Then, the possibilities are
\begin{equation}
\alpha = (\alpha_{ij},\alpha_{jk},\alpha_{ki}), \,
\end{equation}
and in a situation where only two of them are possible, namely the open and closed one, it allows  altogether eight 
three-body channels.  Assuming, two-atom interactions, the potential is an operator that allows transitions between 
the open and closed channels: $  V^{a,b}_{ij}$, which for the moment we are not specifying.

After setting the structure of the Hilbert space where the wave function is defined, we write the three-body Hamiltonian 
as a matrix, with operators as matrix elements in the open-closed channel two-atom states, as
\begin{equation}
  H_{\alpha\beta}= (  H_0 +\Delta_\alpha)\delta_{\alpha\beta}+ \sum_{i>j}   V^{\alpha_{ij},\beta_{ij}}_{ij}
  \delta_{\alpha_{jk},\beta_{jk}}\,\delta_{\alpha_{ki},\beta_{ki}}\, ,
\end{equation}
where, to simplify the presentation, we introduce the following notation
\begin{equation} 
  V^{\alpha\beta}_{ij}:=   V^{\alpha_{ij},\beta_{ij}}_{ij}\, \delta_{\alpha_{jk},\beta_{jk}}\,\delta_{\alpha_{ki},\beta_{ki}}\, ,
\end{equation}
for the potential. The kinetic energy operator is $\hat H_0$ and $\Delta_\alpha$ is the energy level of channel $\alpha$. 
The energy of the channels are given by
\begin{equation}
\Delta_\alpha=\sum_{i>j} e(\alpha_{ij})\, ,
\end{equation}
where $e(a_{ij})$ is the $a_{ij}$ channel energy. For the open channel we set it as $e=0$, therefore the 
three-body open channel continuum has $\Delta=0$. Furthermore, the standard translational invariance 
applies to $  H_{\alpha\beta}$ as long as the two-atom potential depends only on relative coordinates.

The eigenvalue equation for the energy states of the system reads:
{\small \begin{equation}
\sum_{\beta}   H_{\alpha\beta}|\Psi^\beta\rangle=\sum_{\beta} \bigg[(  H_0+\Delta_\alpha)\delta_{\alpha\beta}+ \sum_{i>j}    
V^{\alpha\beta}_{ij}\bigg]|\Psi^\beta\rangle = E|\Psi^\alpha\rangle\, .
\end{equation}
}We write the Faddeev components of the particular case of the bound-state wave function  as~\cite{Fad} ,
\begin{equation}
|\Psi^\alpha\rangle=\frac{1}{E-  H_0-\Delta_{\alpha}+i\varepsilon}  \sum_\beta\sum_{i>j}V^{\alpha\beta}_{ij}|\Psi^\beta\rangle\, ,
\end{equation}
where 
\begin{equation}
G^\alpha_0(E) \equiv \frac{1}{E-  H_0-\Delta_{\alpha}+i\varepsilon}\,
\end{equation}
is the resolvent. The corresponding Faddeev equations can be written as 
\begin{equation}\label{faddeq}
|\Psi^\alpha_i\rangle=G^\alpha_0(E) t^{\alpha\beta}_{jk}(E)\big(|\Psi^\beta_j\rangle+|\Psi^\beta_k\rangle\big)\, ,
\end{equation}
with the two-body T-matrix within the three-body system being a solution of
\begin{equation}
 t^{\alpha\beta}_{jk}(E)=  V^{\alpha\beta}_{jk}+  V^{\alpha\gamma}_{jk}G^\gamma_0(E) t^{\gamma\beta}_{jk}(E)\, .
\end{equation}
For the sake of clarity, when written solely for the two-body subsystem, it reads:
 \begin{equation}
 t^{ab}_{jk}(E)=  V^{ab}_{jk}+  V^{ac}_{jk}\frac{1}{E-H_0^{(2)}-e(c)+i\varepsilon} t^{cb}_{jk}(E)\, ,
\end{equation}
where $H_0^{(2)}$ is the two-body kinetic energy operator. 

Let us consider the most simple AAB case, when only the AA subsystem has open and closed channels, while AB does 
not interact in a closed channel. In such simplified situation the three-body system has only two channels. Then, close 
to a s-wave Feshbach resonance, the  two-body scattering amplitude (evidently associated with the open channel 
$o$ and $a=b=o$) is customarily approximated by the corresponding effective range expansion:

{\small\begin{equation}\label{effrangeexp}
\langle \vec k'| t^{oo}_{jk}(E)|\vec k\rangle=-\frac{1}{(2\pi)^2\mu_{ij}}\left[-a^{-1}+\frac12r_0k^2+\cdots-ik\right]^{-1},
\end{equation}
}where $k=\sqrt{2\mu_{ij}E}$,  $a_0$ is the scattering length and $r_0$ the effective range. 

\subsection{Single channel reduction and induced three-atom interaction}\label{3bodyint}

It is possible to reduce the Hamiltonian eigenvalue equation to the open channel, at the expenses of introducing an effective 
Hamiltonian, which can be derived by using the Feshbach projection operators to the open channels ($P$-space) and  closed 
channels  ($Q$-space), respectively, such that $P+Q=1$. In our notation,
\begin{equation}
P_{\alpha\beta}=\delta_{\alpha o} \delta_{o\beta} \quad \text{and} \quad Q_{\alpha\beta}=\delta_{\alpha\beta}-P_{\alpha\beta}\, ,
\end{equation}
where $o$ indicates that the three pair of particles are in two-body open channels. 
However, if we do the common procedure, we loose the re-sum  of  two-body intermediate state propagation  
in the closed channels which builds the two-body T-matrix in the open channel. 
This motivates us to use the Feshbach projection $P$ and $Q$ operators directly in the Faddeev equations. 
We can exemplify the procedure by considering the bound-state in (\ref{faddeq}), such that
\begin{eqnarray}\label{faddeq-1}
|\Psi^\alpha_i\rangle&=&G^\alpha_0(E) t^{\alpha\beta}_{jk}(E)(P_{\beta\gamma}+Q_{\beta\gamma}) 
\big(|\Psi^\gamma_j\rangle+|\Psi^\gamma_k\rangle\big)
\nonumber \\&=&
G^\alpha_0(E) t^{\alpha o}_{jk}(E) \big(|\Psi^o_j\rangle+|\Psi^o_k\rangle\big)
+G^\alpha_0(E) t^{\alpha \beta}_{jk}(E) \big(Q_{\beta\gamma}|\Psi^\gamma_j\rangle+
Q_{\beta\gamma}|\Psi^\gamma_k\rangle\big) .
\end{eqnarray}
For the open channel we have that
\begin{eqnarray}\label{faddeq-2}
|\Psi^o_i\rangle&=&G^o_0(E) t^{o o}_{jk}(E) \big(|\Psi^o_j\rangle+
|\Psi^o_k\rangle\big)
+ G^o_0(E) t^{o \beta}_{jk}(E) \big(Q_{\beta\gamma}|\Psi^\gamma_j\rangle+
Q_{\beta\gamma}|\Psi^\gamma_k\rangle\big). 
\end{eqnarray}
For the closed channels, the Faddeev equations can be written as an inhomogeneous integral equations, given by 
{\small \begin{eqnarray}\label{faddeq-3}
Q_{\alpha\beta}|\Psi^\beta_i\rangle&=&Q_{\alpha\beta}G^\beta_0(E) t^{\beta o}_{jk}(E) \big(|\Psi^o_j\rangle+
|\Psi^o_k\rangle\big)
+ Q_{\alpha\delta}G^\delta_0(E) t^{\delta \beta}_{jk}(E) \big(Q_{\beta\gamma}|\Psi^\gamma_j\rangle+
Q_{\beta\gamma}|\Psi^\gamma_k\rangle\big).
\end{eqnarray}
}
Due to the elimination of the closed channels to describe the dynamics of the system only using the open 
channel, the 
interactions in the Faddeev equations  for the open channel 
gain new terms coming from the virtual propagation of the  three atoms
in a closed channel. 

Such interactions are given automatically by three-particle connected operators, that
means an induced three-atom interaction. To illustrate this, we introduce 
the iterative solution corresponding to Eq.~(\ref{faddeq-3}), given by  
\begin{eqnarray}\label{faddeq-4}
Q_{\alpha\beta}|\Psi^\beta_i\rangle=Q_{\alpha\beta}G^\beta_0(E) t^{\beta o}_{jk}(E) 
\big(|\Psi^o_j\rangle+|\Psi^o_k\rangle\big)+\cdots , 
\end{eqnarray} 
in Eq.~(\ref{faddeq-2}), which results in the single 
channel Faddeev equations:
\begin{small}
\begin{eqnarray}\label{faddeq-5}
|\Psi^o_i\rangle&=&G^o_0(E) t^{o o}_{jk}(E) \big(|\Psi^o_j\rangle +|\Psi^o_k\rangle\big)
+ G^o_0(E) t^{o \beta}_{jk}(E)Q_{\beta\gamma}G^\gamma_0(E) t^{\gamma o}_{ki}(E) 
\big(|\Psi^o_i\rangle+|\Psi^o_k\rangle\big)
\nonumber \\ 
&+&G^o_0(E) t^{o \beta}_{jk}(E)Q_{\beta\gamma}G^\gamma_0(E)t^{\gamma o}_{ij}(E)
\big(|\Psi^o_i\rangle+|\Psi^o_j\rangle\big)\big)+\cdots .
\end{eqnarray}
\end{small}
The new connected operators, contributing to the kernel of the first terms of 
Eq.~(\ref{faddeq-5})  are 
\begin{eqnarray}\label{faddeq-6}
&&t^{o \beta}_{jk}(E)Q_{\beta\gamma}G^\gamma_0(E) t^{\gamma o}_{ki}(E)
\;\; {\rm and}\;\; t^{o \beta}_{jk}(E)Q_{\beta\gamma}G^\gamma_0(E)t^{\gamma o}_{ij}(E)\, ,
\end{eqnarray}
these operators clearly indicate the propagation in closed channels, which characterizes the nature of a 
three-body force as sketched in Ref.~\cite{YamaEPL2006}. 

It is not a complex exercise to extrapolate the above result to the full series of connected kernel, 
considering any number of intermediate propagation in closed channels. 
Therefore, it is clear that the Feshbach resonance also drives a triatomic interaction, 
besides the two-atom scattering length, and potentially could be controlled, at least from 
the theoretical point-of-view. The framework developed so far can also be generalized to four atoms 
with the corresponding Faddeev-Yakubovski equations \cite{Yak}.

\subsection{Effective three-body interaction and  vanishing scattering length}\label{e3inta0}

In atomic traps, the scattering length can be tuned by Feshbach resonance techniques, as shown in Ref.~\cite{1998Inouye}, such that
one can reach a particular situation is which the two-body scattering length is exactly zero. In this limit, the dynamics of the   
Bose-Einstein condensate, as described by the Gross-Pitaesvki equation, says that the interaction between identical bosonic  
atoms ceases to exist. However, as discussed in Sect.~\ref{3bodyint}, in the single channel Faddeev equation we have 
the contribution of an effective three-body force when the closed channels are eliminated in favor of the open channel. 

By assuming that the s-wave two-atom T-matrix in the open channel is obtained for a short-range interaction, close to a 
zero-range form, its matrix element will not depend on the relative momentum, except for the
dependence on the energy of the system. In this case, the matrix elements will resemble the amplitude
written in Eq.~(\ref{effrangeexp}), such that in the $a_0\to 0$ limit, the two-body T-matrix in the 
open channel vanishes, i.e. $t^{oo}_{ij}(E)=0$, and for the identical bosons system, the coupled open-closed 
channels Faddeev equation  (\ref{faddeq-2}) will simplify and given by 

\begin{equation}\label{faddeq-7}
|\Psi^o_i\rangle=G^o_0(E) t^{o \beta}_{jk}(E) \big(Q_{\beta\gamma}|\Psi^\gamma_j\rangle+
Q_{\beta\gamma}|\Psi^\gamma_k\rangle\big)
\, ,
\end{equation}
where the term carrying the open-channel two-body T-matrix disappear, allowing to determine 
the open channel Faddeev components of the wave function  from  the ones in the closed channel.

By eliminating the open channel components in the closed-channel Faddeev equations, (\ref{faddeq-1}),  
by using Eq. (\ref{faddeq-7}) written above, one gets the following set of coupled equations in the closed 
channels:

{\small\begin{eqnarray}\label{faddeq-8}
Q_{\alpha\beta}|\Psi^\beta_i\rangle&=&Q_{\alpha\beta}G^\beta_0(E) t^{\beta o}_{jk}(E)
G^o_0(E) t^{o \gamma}_{ki}(E) \big(Q_{\gamma\delta}|\Psi^\delta_k\rangle+Q_{\gamma\delta}|\Psi^\delta_i\rangle\big)
\nonumber \\
&+&Q_{\alpha\beta}G^\beta_0(E) t^{\beta o}_{jk}(E)
G^o_0(E) t^{o \gamma}_{ij}(E) \big(Q_{\gamma\delta}|\Psi^\delta_i\rangle+Q_{\gamma\delta}|\Psi^\delta_j\rangle\big)
\nonumber \\
 &+&Q_{\alpha\delta}G^\delta_0(E) t^{\delta \beta}_{jk}(E) 
 \big(Q_{\beta\gamma}|\Psi^\gamma_j\rangle+Q_{\beta\gamma}|\Psi^\gamma_k\rangle\big).
\end{eqnarray}
}
The connected operators in the first two terms of the equation, which can be interpreted as an effective three-body 
interaction, come from the coupling of the closed and open channels with a virtual propagation of the system 
to the open channel coming back to the closed one. This is can be seen, for example, in the first term of the kernel
 \begin{equation}
 Q_{\alpha\beta}G^\beta_0(E) t^{\beta o}_{jk}(E)G^o_0(E) t^{o \gamma}_{ki}(E) Q_{\gamma\delta}.
 \end{equation}
The indices $\beta$ and $\gamma$ in 
 $ t^{\beta o}_{jk}(E)$ $G^o_0(E) t^{o \gamma}_{ki}(E)$ are referring to the closed channels 
and  $G^o_0(E)$ corresponds to the virtual propagation of the three-body system in the open channel.  

This extreme situation can be studied in schematic models in the limit of  zero-range interactions, for example. 
In what follows we will exemplify another case modeled with a two-channel s-wave zero-range interaction model, 
close to the Feshbach resonance. This setup allows the derivation of an analytically two-body T-matrix and the 
built of the bound-state Faddeev equations.

\section{ AAB system  with open and closed channels: zero-range model}\label{sec:AAB}

We substantiate the model proposed in Ref.~\cite{YamaEPL2006} by assuming a Feshbach 
resonance in only one pair. The two identical bosons A interact with a third particle B, and 
all s-wave interactions have zero-range. First, we derive the two-body T-matrix for the AA 
system, and then the bound-state Faddeev equations for the open-closed channel AAB model 
are derived in the unitarity limit, $a_0\to\pm\infty$. The parametric dependencies on the 
Feshbach resonance are pointed out. 

\subsection{Two-channel zero range interaction model}\label{sec:t2}

The T-matrix of a two channel zero-range interaction in s-wave can be analytically derived, having a simple form
when the renormalized interaction strength in the open channel vanishes. Such assumption, taken for simplicity, 
allows us to study the three-body dynamics close to the Feshbach resonance, where the coupling with the closed 
channel puts the system in the unitarity limit. 

This model can be renormalized by subtracting the resolvent at some given scale (see e.g. \cite{fredericoppnp2012}), 
which for convenience we choose for zero energy. The two-body T-matrix elements of the Lippmann-Schwinger 
equation can be written, in this model, in an operatorial-matrix notation as  

{\small\begin{eqnarray}\label{LSeq}
T(E)
&=&\begin{bmatrix}
0  & \eta \\
\eta & \lambda
\end{bmatrix}
+\hspace{-0.1cm}\int d^3p\hspace{-0.05cm}
\begin{bmatrix}
0 & \eta g_0(p^2,E-e;0) \\
\eta g_0(p^2,E;0) &\lambda g_0(p^2,E-e;0)
\end{bmatrix}
T(E)\, ,
\\ \nonumber
\end{eqnarray}
}where  the renormalized 
strengths $\lambda$ and $\eta$ refer to the interaction in the closed channel and the
coupling between the open and close channels, respectively. The resolvent, with outgoing boundary condition, is given by 
\begin{equation}
g_0(p^2,E;\mu^2)=\frac{E+\mu^2}{(\mu^2+\frac{p^2}{2m_r} -i\varepsilon)(E-\frac{p^2}{2m_r} +i\varepsilon)} \,,
\end{equation}
where $m_r$ is the reduced mass of the two atoms.
Note that, the resolvent in the closed channel carries the energy $e$, which allows it to open only
when  $E>e$ and when the scattered particles can transitioned from one channel to the other one. 
In what follows, we will study the situation where $E<e$. The operator $T(E)$, in momentum space, 
has matrix elements that do not depend on any momenta, only depending on $E$ - this is a consequence 
of the zero-range interaction, used in this example.

The T-matrix equation \eqref{LSeq} allows analytical solution in the following form, with the 
channel terms given by
\begin{eqnarray}
t^{oo}(E)&=&\left[{\frac{1-\lambda B(E-e)}{\eta^2B(E-e)}-B(E)}\right]^{-1}, \\
t^{oc}(E)&=&t^{co}(E)={\eta}\left[{1-B(E-e)(\eta^2B(E)+\lambda)}\right]^{-1}, \\
t^{cc}(E)&=&\left[{\frac{1}{\eta^2B(E)+\lambda}-B(E-e)}\right]^{-1},
\end{eqnarray}
where
{\small\begin{equation}
B(E)\equiv \int d^3p\frac{E}{(\frac{p^2}{2m_r} -i\varepsilon)(E-\frac{p^2}{2m_r} +i\varepsilon)}
 =-i(2\pi)^2\sqrt{2m_r^3 E}\, ,
\end{equation}
}where we work with the hypothesis of a closed channel, $E<e$, such that
$B(E-e)=(2\pi)^2\sqrt{2m_r^3 (e-E)}$. Therefore, 
\begin{eqnarray}\label{t2}
t^{oo}(E)&=&\left\{4\pi^2m_r \left[{\frac{1}{a_0(E)}+i\sqrt{2m_rE}}\right]\right\}^{-1}\, , \\
t^{oc}(E)&=&t^{co}(E)={\eta}\left[{1-\zeta(E) \sqrt{2m_r (e-E)}}\right]^{-1}\, , \\
t^{cc}(E)&=&\left\{4\pi^2m_r \left[{\frac{1}{\zeta(E)}-\sqrt{2m_r(e-E)}}\right]\right\}^{-1} , 
\end{eqnarray}
where 
\begin{eqnarray}
a_0(E)&\equiv&\frac{(2\pi)^4m^2_r\eta^2\, \sqrt{2m_r (e-E)}}{1-(2\pi)^2 m_r\lambda\,   \sqrt{2m_r( e-E)}},
\\
\zeta(E)&\equiv& 4\pi^2m_r\left[\lambda-i\eta^2  (2\pi)^2m_r\sqrt{2m_rE} \right].
\end{eqnarray}
This example shows that by varying $e$ one can tune $a_0(0)$, namely the scattering length. 
The model could be enriched by also considering the interaction strength in the open channel. This is left 
for another work. 

One interesting situation that emerges in this model corresponds to the unitarity limit, $a_0(0)\to\pm\infty$. 
In this situation $\lambda^{-1}=(2\pi)^2m_r\sqrt{2m_re}$, giving 

{\small \begin{eqnarray}\label{t2a0}
t^{oo}(E)&=&\left\{4\pi^2m_r \left[{\frac{1}{a_U(E)}+i\sqrt{2m_rE}}\right]\right\}^{-1}, \\
t^{oc}(E)&=&t^{co}(E)=-{i}\left[{2(2\pi)^4m_r^3\eta \sqrt{E (e-E)}}\right]^{-1}\, , \\t^{cc}(E)&=&\frac{1}{(2\pi)^2m_r}\left[\frac{1-i 2 (2\pi)^4\eta^2 m_r^3\sqrt{E(e-E)}}{ 
i\eta^2  2(2\pi)^4m_r^3(e-E)\sqrt{2m_rE}}\right] ,
\end{eqnarray}
}where 
\begin{eqnarray}
a_U(E)\equiv \frac
{(2\pi)^4m^2_r\eta^2\, \sqrt{2m_r e (e-E)}}{\sqrt{e}-  \sqrt{ e-E}}\,.
\end{eqnarray}
For $E<0$, which corresponds to the three-body bound state,  
{\small\begin{equation}
t^{cc}(E)=-\frac{1}{(2\pi)^2m_r}\left[\frac{1+\eta^2  2(2\pi)^4m_r^3\sqrt{|E|(e-E)}}{ 
\eta^2  2(2\pi)^4 m_r^3(e-E)\sqrt{2m_r|E|}}\right] ,
\end{equation}
}leading to a strong attraction in the closed channel within the kernel of the Faddeev equations. The other term  
$t^{oc}(E)$, from the coupling between the open and closed channels, has an attractive effect in 
three atom system. This will be discussed in the next section.

\subsection{AAB coupled channel  model in the unitarity limit}\label{sec:AABun}

Our starting point is Eq. \eqref{faddeq} and the two-body T-matrix \eqref{t2a0} 
for $a_0=\pm\infty$. We are assuming that the subsystem AA has two channels 
and $AB$ has only one open channel. 
The first observation is that the matrix elements of the T-matrix for the AA system, derived in 
Sect. \ref{sec:t2} in momentum space, depends only on the energy, as a consequence of the 
of the zero-range interaction s-wave. This is also the case for the single channel AB T-matrix. 
For this particular AAB system, in units of $m_A=m_B=1$  and $\hbar=1$, it follows that

\begin{equation}\label{spec}
\langle \vec q_{A(B)}\vec p_{A(B)} |\Psi^{\alpha}_{A (B)}\rangle=\frac{f^{\alpha}_{A(B)}
(\vec q_{A(B)})}{E-\frac34 q_{A(B)}^2-p_{A(B)}^2-\Delta_\alpha}\, ,
\end{equation}
where $\vec q_{A(B)}$ is the relative momentum of particle $A(B)$ with respect to the center 
of mass of the pair $A' B (AA)$, with $\vec p_{A(B)}$ the relative momentum of the particles 
$A' $ and $B$ ($A$ and $A'$).

Considering that the bosonic wave function is symmetric by the exchange of the 
identical bosons A and A', it follows that the spectator functions $f^\alpha_i$ should 
be:
\begin{equation}
f^\alpha_A(\vec q_A)=f^\alpha_{A'}(\vec q_{A'}) \equiv f^\alpha_A(\vec q)\,,
\end{equation}
for $\vec q_A=\vec q_{A'}\equiv\vec q$. For the AAB system only two channels are present: 
AA standing in the open or closed channels. The correspondent spectator 
functions are $f^\alpha$, with $\alpha=0,1$, respectively, for the open, (o), 
and closed, (c), channels, indicating the pair AA propagating in 
the open or closed channels, and correspondingly $\Delta_0=\alpha\, e $. 
Altogether for this model, the spectator functions are four:
$ f^0_A(\vec q)\,, ~  f^1_A(\vec q)\, ,~  f^0_B(\vec q)\, $ and $ f^1_B(\vec q)$.  

In the unitary limit, i.e., by assuming $a_{AB}=a_{AA}\to \pm \infty$ 
for both interactions,  the set of four coupled Faddeev equations are derived from  
Eq.~\eqref{faddeq} by using Eq.~\eqref{spec}: 

{\small
\begin{eqnarray}\label{f0A}
&&f^0_A(\vec q)=
 \frac{1}{2\pi^{2}\sqrt{|E|+\frac34q^2}}\int d^3k \, {\cal K}_E^0(\vec q,\vec k)
 \bigg(f^0_A(\vec k)+f^0_B(\vec k)\bigg),
\\ 
\label{f0B}
&&f^0_B(\vec q)=
\frac{1/(2\pi^4\eta)}{\sqrt{\left(|E|+e+\frac34 q^2\right)\left(|E|+\frac34 q^2\right)}}\int d^3k \, 
{\cal K}^1_E(\vec q,\vec k)f^1_A(\vec k) \nonumber \\
 &&+\frac{1/\pi^2}{\left[\frac{-1}{a_U(-|E|-\frac34q^2)}+\sqrt{|E|+\frac34q^2}\,\right]}\int d^3k \, 
 {\cal K}_E^0(\vec q,\vec k)f^0_A(\vec k) 
 \, , \\ 
  \label{f1A}
&&f^1_A(\vec q)=
 \frac{1/(2\pi^{2})}{\sqrt{|E|+e+\frac34q^2}}\int d^3k \, {\cal K}_E^1(\vec q,\vec k)
 \bigg(f^1_A(\vec k)+f^1_B(\vec k)\bigg), \\ 
   \label{f1B}
&& f^1_B(\vec q)=\frac{1/(2\pi^4\eta)}{\sqrt{\left(|E|+e+\frac34q^2\right)\left(|E|+\frac34 q^2\right)}  }\int d^3k \,
 {\cal K}^0_E(\vec q,\vec k)f^0_A(\vec k)
   \\
&&+ \frac{{1}/{(2\pi^2\eta)^2}+\sqrt{\left(|E|+e+\frac34q^2\right)\left(|E|+\frac34 q^2\right)}}{ 
  \pi^2 (|E|+e+\frac34q^2)\,\sqrt{|E|+\frac34 q^2}}\int d^3k \, {\cal K}^1_E(\vec q,\vec k)f^1_A(\vec k) 
 \, ,  
  \nonumber  
 \end{eqnarray}
}
where the three-body resolvent in terms of the spectator momenta is given by:
\begin{equation}\label{kernel}
{\cal K}^\alpha_E(\vec q,\vec k)=\frac{1}{|E|+\alpha \, e+ q^2+k^2+\vec q .\vec k}\, ,
\end{equation}
which takes into account the energy gap between the two channels.

The coupled set of equations \eqref{f0A} and \eqref{f0B} for the spectator 
functions $f^0(\vec q)$  and $f^0_B(\vec q)$ reduces to the Skorniakov and 
Ter-Martirosian equations~\cite{STM} for an AAB system, which for three atoms in s-wave 
present both to the Efimov effect and to the Thomas collapse.
The Efimov effect and Thomas collapse are related to the breaking of 
continuous scale invariance to a discrete one. This demands the necessity of an 
ultraviolet (UV) scale associated to the three-body one, which carries 
all correlations of physical observables of the s-wave three-particle system. 
These correlations can easily be represented by scaling functions 
(see \cite{fredericoppnp2012}).

We have discussed in Sect. \ref{3bodyint} the appearance of an effective three-body interaction, when the
Faddeev equations in the open and closed channels are reduced to a single one.  The set of coupled integral equations 
for the spectator functions of the AAB system, \eqref{f0A}-\eqref{f1B}, illustrates such dynamical mechanism.
The coupling with $f^1_A$ in the open channel equation \eqref{f0B} could be translated to an effective three-body 
interaction acting in the open channel equations for $f^0_A(\vec q)$  and $f^0_B(\vec q)$. 

The coupled set of equations  \eqref{f0A}-\eqref{f1B} needs to be regularized at the UV region to avoid the
collapse of the AAB system for vanishing total angular momentum. The regularization can be done by 
resorting to a subtraction in the kernel. Then instead of \eqref{kernel} we can use the subtracted 
resolvent~\cite{AdhPRL1995}:

{\small\begin{eqnarray}\label{kernelsub}
{\cal K}^\alpha_E(\vec q,\vec k)&=&\frac{1}{|E|+\alpha \, e+ q^2+k^2+\vec q .\vec k}
-\frac{1}{\mu^2+\alpha \, e+ q^2+k^2+\vec q .\vec k}\, 
 \nonumber \\ \\
&=&\frac{\mu^2-E}{\left(|E|+\alpha \, e+ q^2+k^2+\vec q .\vec k\right)\left(\mu^2+\alpha \, e+ q^2+k^2+\vec q .\vec k\right)}\, , 
\nonumber \end{eqnarray}
}such that $\mu^2>>e$.  It will imply that, even with $\mu^2$ fixed, the effect of the coupling 
between the open and closed channels, even at the unitarity limit,  
allows to drive the Efimov states  by changing $\eta$ and/or $e$.  
In this model, the relevant dimensionless quantities for driving the Efimov states at 
the unitarity are $e/\mu$ and $\eta\, \mu$. The energy of the AAB system is given 
in units of $\mu$. 

The short-range scale is related to the van der Waals interaction as $\mu\sim1/\ell_{vdW}$. At unitarity, the
single channel description of the AAB system makes that all s-wave three-body observables be scaled 
with powers of $\mu$. This is a direct consequence of the possibility to find the particles simultaneously 
at the same position, i.e., depending on the configuration of the constituents of the system, this collapse 
can exist for bosonic or fermionic systems. For example, in the single channel framework 
given by Eqs. \eqref{f0A} and \eqref{f0B}, with the subtracted kernel \eqref{kernelsub}, once 
the coupling with the closed channel is disregarded, one can easily verify 
that at the unitarity the three-body binding energy is proportional to $\mu^2$, as no 
other scales are present in the integral equations in limit of a zero-range interaction. 
The separation and coupling between the open and closed channel should represent other 
length scales, namely $\eta$ and $1/\sqrt{e}$, which controls the scattering length 
for $E=p^2<e$
\begin{eqnarray}
p\cot\delta^U_0 &=&-\frac{\sqrt{e}-  \sqrt{e-p^2}}
{4\pi^4\eta^2\,\sqrt{e}\, \sqrt{e-p^2}}\Bigg|_{p^2\to 0}
= - \frac12R^*p^2 \,\, \text{with} \,\,\,R^*=\frac{1}{4\pi^4\eta^2  e\sqrt{e}}\,,
\end{eqnarray}
where the parameter $R^*$ is controlled by the length scale  $\gamma=\eta^{-2}e^{-\frac32}$,  
with $R^*$ related to the width of the Feshbach resonance~\cite{2004Petrov}.
 
The separation energy $e$ between the two channels gives the momentum scale $\sqrt{e}$, 
which competes in the UV region with the subtraction scale, 
as it is seen in the coupling term between the open and closed channels in Eq.  \eqref{f0B}. Note that 
for momentum $k\sim\sqrt{e}$, the virtual propagation of the system brings to the bound state a new
scale in the UV region competing with $\mu^2$. 
The van der Waals radius is associate to an equivalent 
energy scale of $\hbar^2/(m\ell_{vdW}^2)$. For example, in the  $^{133}$Cs$_2$ system, where 
$\ell_{vdW}=101\,a_0$~\cite{2010Chin} ($a_0$ the Bohr radius) the energy
 $\mu^2\,\sim\, 127~\mu$K sets the UV scale for the $^{133}$Cs$_3$ system. 
The energy gap between the open and closed channels will compete, in the UV region, with the subtraction 
scale that can drive the triatomic system, beyond the scattering and effective range, by moving the 
Feshbach resonance parameters.

\section{Final considerations}\label{sec:final}

In summary, the three-atom bound state problem with coupled open-closed channels close to a Feshbach resonance is 
formulated  through the Faddeev equations for the wave function. The effective three-body force appearing in the single 
channel description was derived using the Feshbach projection operators in the open and closed channels. Following 
this formal presentation, we investigated the interesting case for which the trap setup is tuned to vanishing scattering 
lengths, such that the direct open channel interaction disappears. In this case, we demonstrated that an effective 
interaction in the open channel appears due to an effective three-body interaction built from the coupling between the 
open and  closed channels. 

We have provided one explicit example of a three-body system composed by two 
identical bosons and a third different particle (AAB system) with open and closed channels, which realizes in practice 
the schematic discussion we have performed in Ref.~\cite{YamaEPL2006}. In this illustrative case, the subsystem AA has 
an open and a closed channel separated by an energy gap, and interacting through a zero range s-wave potential. The 
transition matrix for the AA subsystem is derived resorting to a subtractive renormalization scheme, at the expense 
of introducing two finite parameters, besides the energy gap. The subsystem AA and AB are tuned to the unitary limit, 
and the Faddeev equations for the bound AAB system was derived explicitly, to expose the dependences of  
the kernel with the two body parameters and energy gap. The set of coupled equations allowed to identify 
the effect of  the coupling between the Faddeev components in the open channel with the closed channel ones. The 
energy gap between the channels essentially provides an UV scale probed by the virtual propagation of the three-body 
system in the closed channel. This  short-range scale competes with the van der Waals radius that sets the short-range 
physics of the system in the open channel. Therefore, we have clearly illustrated how the three-atom system can be 
tuned by the Feshbach resonance beyond the scattering length and van der Waals radius.

As a final comment, considering the different possibilities to study few-body physics in atomic traps, we 
should mention that few-atom system can also be driven by the change in the effective dimensions, by squeezing 
the trap, while tuning both the two and few-body parameters controlling the Feshbach resonance. Controlled cold 
chemical reactions~\cite{2008Krems,2009Carr,2010Ospelkaus} with forces driven by external fields is also a field 
of intense activity (see e.g. \cite{2017Rui,2018Hoffmann,2018Gao,2019Yang,2019Li,2019Wang,2020Reynolds,2020Makrides}).

\section*{Acknowledgements}
This article is dedicated to the memory of Prof. Mahir Saleh Hussein, for his relevant contributions in Physics.
We thank Vanderlei Bagnato for calling our attention to the possibility of zero scattering length in atomic traps.
The following Brazilian agencies are being acknowledge for their partial support: 
Conselho Nacional de Desenvolvimento Cient\'\i fico e Tecnol\'ogico
[Procs. 303579/2019-6 (MTY), 308486/2015-3 (TF), 304469-2019-0(LT) and INCT-FNA Proc.464898/2014-5]  and
Funda\c c\~ao de Amparo \`a Pesquisa do Estado de S\~ao Paulo
[Projs. 2019/00153-8 (MTY), 2017/05660-0 (LT and TF) and 2019/07767-1 (TF)].

\end{document}